\documentclass[11pt,a4paper]{article}
\pdfoutput=1
\usepackage{jheppub}
	\usepackage{amsmath, amssymb} 
	\usepackage{amsfonts}
	\usepackage{dsfont}
	\usepackage[ansinew]{inputenc}
	\usepackage[usenames,dvipsnames]{pstricks}
	\usepackage{epsfig}
	\usepackage{pst-grad} 
	\usepackage{pst-plot} 
	\usepackage{hyperref}
	\usepackage{graphicx}
	\usepackage{caption} 
	\usepackage{subcaption}
	\hypersetup
	{
		colorlinks,%
		citecolor=black,%
		linkcolor=black,%
		urlcolor=black,%
	}

\newcommand{\be}{\begin{equation}}
\newcommand{\ee}{\end{equation}}

\def\bC {\mathbb{C}}
\def\bP {\mathbb{P}}

\title{A novel approach to cosmological particle production}

\author{Bartomeu Fiol,}
\author{Mickael Komendyak and}
\author{Jordi Ruiz-Ponsell}

\affiliation{Departament de Física Qu\`antica i Astrof\'isica, Institut de Ci\`encies del Cosmos (ICCUB), \\
Universitat de Barcelona, Mart\'i i Franqu\`es 1, 08028 Barcelona, Catalonia, Spain}

\emailAdd{bfiol@ub.edu}
\emailAdd{komendyak@icc.ub.edu}
\emailAdd{jruizpon12@alumnes.ub.edu}

\abstract{In this work we present a novel approach to the study of cosmological particle production in asymptotically Minkowski spacetimes. We emphasize that it is possible to determine the amount of particle production by focusing on the mathematical properties of the mode function equations, {\it i.e.} their singularities and monodromies, sidestepping the need to solve those equations. We consider in detail creation of scalar and spin 1/2 particles in four dimensional asymptotically Minkowski flat FLRW spacetimes. We explain that when the mode function equation for scalar fields has only regular singular points, the corresponding scale factors are asymptotically Minkowski. For Dirac spin 1/2 fields, the requirement of mode function equations with only regular points is more restrictive, and picks up a subset of the aforementioned scale factors. For the scalar case, we argue that there are two different regimes of particle production; while most of the literature has focused on only one of these regimes, the other regime presents enhanced particle production. On the other hand, for Dirac fermions we find a single regime of particle production. Finally, we very briefly comment on the possibility of studying particle production in spacetimes that don't asymptote to Minkowski, by considering mode function equations with irregular singular points.
}

\begin{document}
\maketitle


\section{Introduction}
Gravitational particle production caused by an expanding universe is one of the earliest and most striking results of
quantum field theory in curved spacetimes, together with Hawking radiation and the study of quantum fluctuations in inflation \cite{Birrell:1982ix, Parker:2009uva}. Particle production in expanding universes was foreseen by Schr\"odinger in a visionary paper \cite{schro}, and was fully established by Parker \cite{Parker:1969au, Parker:1971pt}, see \cite{Kolb:2023ydq, Ford:2021syk} for recent reviews. In a nutshell, the breaking of time translation invariance implies that the energy of the quantum field is not conserved, allowing for particle creation; on the other hand, if the expansion does not break spatial translation invariance, momentum is conserved and single particle creation is not allowed: particles are created in pairs. Besides its intrinsic conceptual interest, the possibility of  purely gravitational particle creation has found many applications in cosmology: from a potential mechanism for the origin of dark matter, to playing a role in baryogenesis \cite{Kolb:2023ydq, Ford:2021syk}. 

The initial discussion \cite{Parker:1969au, Parker:1971pt} and fully solvable examples \cite{parker76, Bernard:1977pq, Duncan:1977fc} considered mostly expanding universes that are asymptotically Minkowski in the distant past and the distant future, where a free quantum field, {\it e.g.} a scalar or a fermionic field, is placed. The choice of universes that asymptote to Minkowski in the asymptotic past and the asymptotic future was motivated by technical simplicity and also by conceptual clarity: in Minkowski spacetime we have an unambiguous definition of vacuum of the quantum theory and of the notion of particle. 

Still in the realm of asymptotically Minkowski spaces, while the existence of fully solvable examples is always welcome, it is also important to elucidate generic properties of cosmological particle production, valid for wide families of expanding universes. A possibility is to develop approximation methods, {\it e.g.} \cite{Zeldovich:1971mw, Birrell:1979pi} to estimate the amount of particle production. In this work we will advocate a different approach, and provide generic results on scalar and spin-1/2 particle production, based on the mathematical properties of the relevant equations, without actually solving them. Let's sketch the main idea and results obtained in this work, focusing for concreteness on the scalar case. We will consider a four-dimensional Friedmann-Lema\^itre-Robertson-Walker (FLRW) flat spacetime, described by a scale factor $a(\eta)$. A scalar field placed in this background can be decomposed in modes $\chi_{\vec k}(\eta)$, that satisfy an equation of the form
\be 
\frac{d^2 \chi_{\vec k}(\eta)}{d\eta^2} +\omega^2 (\vec k,\eta) \, \chi_{\vec k}(\eta)=0,
\label{modeintro}
\ee
where $\omega$ depends on the scale factor $a(\eta)$. Eq. (\ref{modeintro}) is a second order ordinary differential equation (ODE). For our purposes, the relevant characterization of such equations is given by the number and types of singular points it presents (see the appendix for relevant mathematical background). We will first argue that if (\ref{modeintro}) presents only {\it regular} singular points (RSPs), then $a(\eta)$ tends to constant values in the infinite past and future, and thus the spacetime is asymptotically Minkowski. For instance, for a massive conformally coupled scalar, the scale factors $a(\eta)$ fow which the discussion applies are of the form
\be 
a(\eta)^2 =\frac{a_f^2 e^{2 n \eta/s} + \epsilon_{2n-1} e^{(2n-1) \eta/s}+\dots+\epsilon_1 e^{\eta/s}+(c_1\dots c_n)^2 a_i^2}{(e^{\eta/s}+c_1)^2\dots (e^{\eta/s}+c_n)^2},
\label{epsintro}
\ee
where $a_i,a_f$ are the initial and final values of $a(\eta)$, $s$ is a time scale, $n$ is a positive integer, and $\epsilon_i, c_i$ are constants with some positivity constraints.

As is well known, one can introduce two bases of solutions of (\ref{modeintro}), one that tends to plane waves in the asymptotic past, and one that tends to plane waves in the asymptotic future. These two bases are related by a change of basis, given by Bogoliubov coefficientes $\alpha_{\vec k}$ and $\beta_{\vec k}$. If the initial state is the vacuum, cosmological particle production is characterized by a non-zero value of $|\beta_{\vec k}|^2$ \cite{Parker:1969au, Parker:1971pt}.

In this work, we derive a formula for $|\beta_{\vec k}|^2$ for all spacetimes such that the resulting mode function equation (\ref{modeintro}) has only regular singular points. In the case of a massless minimally coupled scalar, it is of the form,
\be 
|\beta_{\vec k}|^2= \frac{\cosh{2\pi s (\omega_f-\omega_i)}-T/2}{\cosh 2\pi s (\omega_f+\omega_i)-\cosh 2\pi s (\omega_f-\omega_i)}
\label{bogointro}
\ee
where $\omega_{i,f}$ are the asymptotic initial and final values of $\omega$ in (\ref{modeintro}), and $T$ is the trace of the {\it composite monodromy} matrix of a path in the Riemann sphere that encircles the two regular singular points in the asymptotic past and the asymptotic future. In general it is possible, but rather elaborate, to write $T$ in terms of the parameters that enter the scale factor (\ref{epsintro}), and there is a large mathematical literature devoted to this problem. We will illustrate the general idea with the simplest examples. In the context of classical reflection of waves in inhomogeneous media, and for the particular case when eq. (\ref{modeintro}) has precisely three singular points a formula similar to (\ref{bogointro}) was derived already in \cite{epstein} (with obviously a different physical interpretation). 

Note that we will derive the amount of particle production (\ref{bogointro}) without actually solving the mode function equation (\ref{modeintro}). By its very nature, the determination of  cosmological particle production necessarily involves non-local information about the mode function equation, beyond the local data around each regular singular point. Traditionally this non-local information is taken to be the {\it connection problem}: solve the equation around the asymptotic past and the asymptotic future, analytically continue a basis of solutions all the way to the other regular singular point, and write them as a linear combination of the basis at that singular point. The novel approach that we advocate in this work trades that non-local question by a different one: find the trace of the composite monodromy
encircling those two singular points, as a function of the scale factor. While the task of relating the trace of the composite monodromy to the scale factor $a(\eta)$ is still non-trivial, the result (\ref{bogointro}) allows us to draw some general lessons. Specifically, the composite monodromy matrix is in $SL(2,\bC)$, and it is known that such matrices belong to different classes, depending of the square of its trace: elliptic for $T^2<4$, parabolic when $T^2=4$ and hyperbolic for $T^2>4$\footnote{There is a further class: {\it loxodromic}, but their trace is not real, so it doesn't appear in the problem at hand.}. Moreover, while the trace in the elliptic case depends on the parameters in a periodic  - and thus bounded - fashion, it displays exponential behavior in the hyperbolic class. Therefore (\ref{bogointro}) immediately reveals that as we vary the shape of the scale factor $a(\eta)$ in (\ref{epsintro}), the regime of particle production can change quite significantly. The well-known examples \cite{Bernard:1977pq, parker76} fall into the elliptic class of the classification above, but as we will show, they belong to families of scale factors $a(\eta)$ that also include examples in the hyperbolic class, and thus display enhanced particle production.

With the advent of inflation and the discovery of the current cosmic accelerated expansion of the Universe, spacetimes that asymptote to Minkowski in the past and in the future are less relevant for cosmology than when the original works on cosmological particle creation appeared \cite{Parker:1969au, Parker:1971pt}. We nevertheless think that our results are intrinsically interesting, and provide a useful warm-up to tackle a similar study of cosmological particle creation in more generic spacetimes. We will comment on the extension of the present work to other asymptotia at the end of this introduction.

The structure of the paper is as follows. In section 2 we discuss production of scalar particles in asymptotically Minkowski spaces. We argue that a wide family of mode function equations with such asymptotia are given by differential equations that present only regular singular points, called Fuchsian equations. For this case, we present a general formula for particle production, uncover the existence of various regimes, and display new fully solved examples. 

In section 3 we present a similar discussion for production of Dirac fermions in asymptotically Minkowski space. The scale factor $a(\eta)$ enters the mode function equation differently that in the scalar case, and the requirement that the mode function equation is Fuchsian selects scale factors of the form
\be
a(\eta)=a_f-\frac{A_1}{e^{\eta/s}+c_1}-\dots \frac{A_n}{e^{\eta/s}+c_n}
\label{fermscaleintro}
\ee
with the constraint $m^2 a^2 (\eta)\pm i m \dot a (\eta) \rightarrow m^2 a_i^2$ as $\eta \rightarrow -\infty$. If we furthermore are interested in expanding universes, the constants in (\ref{fermscaleintro}) have to satisfy further constraints. We similarly find a general formula for the corresponding $|\beta_{\vec k}|^2$. In contrast with the scalar case, we argue that there is just one regime of particle production.

This work opens various venues for further research. First of all, we have restricted our analysis to scalar and spin 1/2 Dirac particles. It may be interesting to study creation of particles of higher spin in these backgrounds; a first guess is that there will be different regimes of bosonic particle creation, but a single regime for fermions. Also, we have only considered free fields, and it is worth exploring the possibility of extending our results to interacting quantum fields.

A potentially more interesting extension is to apply the approach advocated here to spacetimes that do not asymptote to Minkowski in the past and/or the future. Let's end the introduction by presenting some preliminary ideas in this direction \cite{wop}. In the current work, we have shown that if the mode equation (\ref{modeintro}) is Fuchsian, then generically the scale factor $a(t)$ asymptotes to Minkowski in the past and in the future, eq. (\ref{epsintro}). This immediately suggests that to analyze spacetimes that don't asymptote to Minkowski, we turn our attention to mode equations (\ref{modeintro}) that present irregular singular points in the asymptotic past and/or future. A simple way to do that is to consider a mode equation with only regular singular points, and take the limit when two regular singular points merge, and become an irregular singular point\footnote{The resulting differential equations are known as {\it confluent} in the mathematical literature.}. In the simplest relevant case, consider the Riemann equation with three RSPs at $0,t,\infty$. When we take the limit $t\to \infty$, these two RSPs merge, becoming an irregular singular point; the resulting equation is the Kummer equation. Interestingly enough, the mode equation for pure de Sitter space is a Bessel equation \cite{Birrell:1982ix, Parker:2009uva}, which is a particular case of the Kummer equation. A second interesting example involves a Fuchsian equation with four regular singular points (known as the Heun equation); taking the limit where the four RSPs merge in pairs (the so called doubly confluent Heun equation), we arrive at a profile that doesn't asymptote to Minkowski space-time neither in the past nor in the future. It seems worth to further elucidate the correspondence between irregular singular points of the mode function equation, the asymptotic behavior of the scale factor $a(t)$, and properly defined particle production in those spacetimes\footnote{See \cite{Enomoto:2020xlf} for work on these topics, with a philosophy similar to the one advocated here.}.

\section{Scalar particle production}
The spacetimes we are going to discuss in this work are four-dimensional spatially flat Friedman-Lema\^itre-Robertson-Walker (FLRW) spacetimes. It will be convenient to introduce various time coordinates,
\be
ds^2=-dt^2+ a(t)^2 d\vec x^2 \, = \, a(\eta)^2 (-d\eta^2+d\vec x^2) \, = \, -a(\tau)^6 d\tau^2 +a(\tau)^2 d\vec x^2 .
\label{flrw}
\ee
In this work we restrict to asymptotically Minkowski spacetimes, meaning that we require that $a(t)$ asymptotes to constant values as $t\rightarrow \pm \infty$,
\be
a(t) \rightarrow
\begin{cases}
a_f, & t\rightarrow \infty \\
a_i, & t\rightarrow -\infty
\end{cases}
\ee
On these spacetimes we place a scalar field $\phi(x)$ of mass $m$ and coupling $\xi$ to the Ricci scalar curvature. The equation of motion for this field is
\be
\Box \phi -m^2\phi -\xi R \phi=0.
\label{scalareom}
\ee
Any solution to this equation can be decomposed into a sum over a complete set of positive and negative norm solutions,
\be 
\phi=\sum_{\vec k} \left( a_{\vec k} f_{\vec k}+a^\dagger_{\vec k} f^*_{\vec k}\right).
\ee
The decomposition can be taken to be a discrete sum by imposing periodic boundary conditions \cite{Birrell:1982ix, Parker:2009uva}. In the canonical quantization of the field theory, the coefficients $a^\dagger_{\vec k}, a_{\vec k}$ are promoted to creation and annihilation operators.

For the sake of concreteness, we will discuss two particular examples of (\ref{scalareom}). The first one is a massive field with conformal coupling $\xi =\frac{1}{6}$, and the second one is a minimally coupled $\xi=0$ massless field $m=0$. This second example is relevant for cosmological graviton production \cite{lifshitz, Grishchuk:1974ny,Ford:1977dj}. In the first case, if we write
\be
f_{\vec k} (\eta, \vec x)= \frac{e^{i\vec k \cdot \vec x} \chi_{\vec k} (\eta)}{a(\eta)\sqrt{V}}
\label{confmode}
\ee
with $\eta$ defined in (\ref{flrw}), the equation of motion reduces to 
\be
\frac{d^2 \chi_{\vec k}(\eta)}{d \eta^2} +\left(\vec k^2 + m^2 a(\eta)^2 \right) \chi_{\vec k} (\eta)=0.
\ee
Similarly, for the $m=\xi=0$ case, we can introduce the modes \cite{parker76}
\be
f_{\vec k} (\tau, \vec x)= \frac{e^{i\vec k \cdot \vec x}}{\sqrt{V}} \chi_{\vec k} (\tau)
\label{minmode}
\ee
and the resulting equation of motion is
\be
\frac{d^2 \chi_{\vec k} (\tau)}{d \tau^2} +\vec k^2 \, a(\tau)^4 \, \chi_{\vec k} (\tau)=0.
\ee
In both cases, the equation of motion is a second order ordinary differential equation of the form
\be
\frac{d^2 \chi_{\vec k}(t)}{dt^2}+ \omega (t,|\vec k|)^2 \, \chi_{\vec k}(t)=0.
\label{qform}
\ee
where $t$ stands for either $\eta$ or $\tau$ and
\be
\omega=
\begin{cases}
\sqrt{|\vec k|^2 +m^2 a(\eta)^2}, & \,\, m\neq 0, \,\, \xi =\frac{1}{6}. \\
|\vec k| a(\tau)^2,  & \,\, m=0, \,\, \xi=0.
\end{cases}
\ee
As the spacetime is asymptotically Minkowski in the past and in the future, there will be solutions approaching plane waves as
\be 
\chi_{\text{in},\vec k} \sim \frac{e^{-i\omega_i t}}{\sqrt{2\omega_i}}, \hspace{1cm} \chi_{\text{in},\vec k}^* \sim \frac{e^{i\omega_i t}}{\sqrt{2\omega_i}}, \hspace{1cm} t\rightarrow -\infty.
\label{planepast}
\ee
Note that we have identified these two solutions as complex conjugate of each other. In general, since the ODE (\ref{qform}) has real coefficients, if $\chi$ is a solution, so is $\chi^*$. In particular, if $\chi$ asymptotes to a positive energy plane wave, $\chi^*$ asymptotes to a negative one. Similarly
\be 
\chi_{\text{out},\vec k} \sim \frac{e^{-i\omega_f t}}{\sqrt{2\omega_f}}, \hspace{1cm} \chi_{\text{out},\vec k}^* \sim \frac{e^{i\omega_f t}}{\sqrt{2\omega_f}}, \hspace{1cm} t\rightarrow \infty.
\label{planefut}
\ee
Since the space of solutions of (\ref{qform}) is two-dimensional, it must be possible to express any pair of solutions as a linear combination of any other pair of linearly independent solutions. In particular
\be
\begin{split}
\chi_{\text{in},\vec k}= \alpha_k \, \chi_{\text{out},\vec k} +\beta_k \, \chi_{\text{out},\vec k}^* \, , \\
\chi_{\text{in},\vec k}^*= \beta_k^* \, \chi_{\text{out},\vec k} +\alpha_k^* \, \chi_{\text{out},\vec k}^* \, .\\
\label{thebogo}
\end{split}
\ee
where $\alpha_k,\beta_k$ are the Bogoliubov coefficients. The creation and annihilation operators in both regions are also related by the same Bogoliubov coefficients \cite{Birrell:1982ix, Parker:2009uva}. In particular, if we grant that the initial state was the in-vacuum, $|0\rangle_{\text{in}}$, and write $\phi=\sum \left( b_{\vec k} f^{out}_{\vec k}+b^\dagger_{\vec k} f^{out\, *}_{\vec k}\right)$, the mean number of particles created with mode $\vec k$ is \cite{Birrell:1982ix, Parker:2009uva} 
\be 
\langle N_{\vec k} \rangle = { }_{\text{in}}\langle 0|b_k^\dagger b_k|0\rangle_{\text{in}} =|\beta_{\vec k}|^2 .
\label{betasq}
\ee
Notice that to compute particle production, knowledge of the Bogoliubov coefficients is more important than that of the explicit solutions to (\ref{qform}).   

\subsection{Asymptotic Minkowski spacetimes and Fuchsian equations}
We have seen that the mode function equation (\ref{qform}) is a second order ordinary differential equation that depends on the scale factor $a(t)$. Instead of trying to solve it on a case by case basis, we want to argue that if we restrict the types of singularities that (\ref{qform}) can display, we can draw general conclusions about scalar particle production in those expanding universes, without solving (\ref{qform}).

We will present three general results: first, we will show that if (\ref{qform}) is a Fuchsian equation, {\it i.e.} it presents only regular singular points\footnote{See the appendix for a brief overview of the necessary mathematical background.}, the scale factor asymptotes to constant values in the past and in the future. Then, we will show that the Frobenius basis of solutions discussed in the mathematical literature is essentially (up to normalization) the basis of solutions that asymptote to plane waves, which is the physically relevant basis. This identification implies that the connection formulae for the Frobenius bases immediately gives the relevant Bogoliubov coefficients, eq. (\ref{thebogo}). Finally, we will derive a formula for $|\beta_{\vec k}|^2$ where all the intricacies of the profiles are encapsulated in a single function, and explain that on very generic grounds, there are two regimes for scalar particle production.

Our starting point is the most general second order Fuchsian ODE, that can be written as \cite{haraoka}
\be 
\frac{d^2 \Psi(z)}{dz^2}+\left(\sum_{k=1}^n \frac{A_k}{z-z_k}\right) \frac{d\Psi(z)}{dz}+
\sum_{k=1}^n \left( \frac{B_k}{(z-z_k)^2}+\frac{C_k}{z-z_k} \right) \Psi(z)=0
\label{abcfuchs}
\ee
where $A_k,B_k,C_k$ are constants and $\sum_{k=1}^n C_k=0$. It has regular singular points at $z=z_1,\dots,z_n$, and at $z=\infty$ (except for some values of the coefficients, see appendix). Our goal is to rewrite this equation in the form of (\ref{qform}). As reviewed in the appendix, there are two generic transformations that we can perform on Fuchsian equations: M\"obius and index transformations. In the case at hand, first change variables $z\rightarrow z+z_1$ (this is a particular case of a M\"obius transformation). Next, perform an index transformation, setting $A_1=1, A_2=\dots =A_n=0$. This brings the Fuchsian equation (\ref{abcfuchs}) to the form
\be
\frac{d^2 \chi(z)}{d z^2}+\frac{1}{z} \frac{d \chi(z)}{dz}+ \frac{P_{2n-2}(z)}{z^2 (z-z_2)^2 \dots (z-z_n)^2}\chi(z)=0
\label{abcfuchs2}
\ee
where $P_{2n-2}(z)$ is a degree 2n-2 polynomial\footnote{The reason why it is not of degree $2n-1$ is the $\sum C_k=0$ condition.}. Finally, change variables $z=e^{t/s}$ to arrive at the normal form of the equation
\be
\frac{d^2\chi(t)}{dt^2}+\frac{1}{s^2}\frac{P_{2n-2}(e^{t/s})}{(e^{t/s}-z_2)^2 \dots (e^{t/s}-z_n)^2}\chi(t)=0 .
\ee 
To have a continuous profile, we don't allow any of the $z_i$ to be in the positive real axis. Except for this restriction, we have showed that generic Fuchsian ODEs can be written as (\ref{qform}), where the frequency is of the form
\be
\omega^2(t)=\frac{\omega_f^2 e^{2(n-1)\frac{t}{s}}+\dots+(c_1 c_2\dots c_{n-1})^2 \omega_i^2}{(e^{\frac{t}{s}}+c_1)^2 \dots (e^{\frac{t}{s}}+c_{n-1})^2}
\label{normalfuchsian}
\ee
and we have already identified two of the coefficients of the polynomial in the numerator. It is manifest from (\ref{normalfuchsian}) that as $t\rightarrow +\infty$, $\omega (t)^2 \rightarrow \omega_f^2$ and as $t\rightarrow -\infty$, $\omega (t)^2 \rightarrow \omega_i^2$. So indeed, if the equation for the modes is Fuchsian, it provides a scale factor that asymptotes to Minkwoski in the past and in the future.

In the construction above, $t\rightarrow -\infty$ and $t\rightarrow +\infty$ correspond to regular singular points of the equation of motion. This implies that there are no analytic solutions that can be constructed as power series with integer powers.  The Frobenius method provides a basis of solutions near each regular singular point of the equation. They are of the form
\be
\phi_{\pm}(u)=u^{r_\pm} \sum_k c_k u^k
\label{frobenius}
\ee
where $u$ is a coordinate that vanishes at the RSP, and $r_\pm$ are the two roots of the {\it indicial equation} associated to the RSP, see the appendix for details. From (\ref{normalfuchsian}) it is immediate that the respective indicial equations to these RSPs are,
\be
r^2+s^2 \omega_i^2=0, \hspace{1cm} r^2+s^2 \omega_f^2=0.
\label{indicial}
\ee
At the z=0 RSP we take u=z, at the $z=\infty$ RSP, we take $u=1/z$. Then, the change of variables $z=e^{t/s}$, shows that the respective asymptotic behaviors of the Frobenius solutions (\ref{frobenius}) are $e^{\pm i \omega_i t}$ and $e^{\mp i \omega_f t}$. These are precisely the asymptotic behaviors of the plane waves, except for the fact that the plane wave solutions are normalized  as in (\ref{planepast}) and (\ref{planefut}).


Once we have identified these two bases of solutions, we are ready to discuss the change of basis that relates them. In the context of differential equations, finding the linear transformation that relates solutions defined as power series around different RSPs is referred to  as the {\it connection problem} \cite{haraoka}. Finding the explicit form of the connection matrix between the Frobenius bases of two RSPs is rather complicated. However, in the case at hand there are two features that drastically simplify the discussion. First, as emphasized above, in our case the two solutions in each basis are complex conjugate to each other, and that implies relations among the components of the connection matrix (\ref{thebogo}). Second, it follows from (\ref{betasq}) that we don't need to know every component of the connection matrix (\ref{thebogo}), just $|\beta_{\vec k}|^2$. These simplifications allow us to find a generic formula for $|\beta_{\vec k}|^2$ that depends very simply on a single function that encodes of the details of the profile.

First, if we work with the plane wave normalization for the solutions, the connection matrix is an element of $SL(2,\bC)$. In the particular case we are dealing with, the two basis elements are complex conjugate of each other, eq. (\ref{thebogo}), and that implies the connection matrix is symplectic,
\be
|\alpha_{\vec k}|^2-|\beta_{\vec k}|^2=1.
\label{bogodet}
\ee

To proceed, the key idea is to think of $z$ in eqs. (\ref{abcfuchs}) and (\ref{abcfuchs2}) as a complex variable, and consider elementary and composite monodromies for the two relevant RSPs in the $z$ Riemann sphere. For the elementary monodromies around each of the two RSPs, the monodromy matrices in the respective Frobenius bases are of the form,
\be
\begin{pmatrix}
e^{2\pi i r_+} & 0 \\
0 & e^{2\pi i r_- } 
\end{pmatrix}
\ee
where $r_\pm$ are the roots of the indicial equations, eq. (\ref{indicial}). Note that these two monodromy matrices are in $SL(2,\bC)$. The composite monodromy corresponds to a path that encircles the two RSPs at the asymptotic past and asymptotic future (and no more than those two singular points). Since the matrices for elementary monodromies are in $SL(2,\bC)$, the composite monodromy matrix ${\cal M}_{0\infty}$ is also in $SL(2,\bC)$. In particular, it can be brought to upper triangular form, with the diagonal components being $\{-e^{2\pi i \sigma}, -e^{-2\pi i \sigma}\}$, where $\sigma$ will be in general complex, and the minus sign in front of the exponents is a convenient convention. $\sigma$ is a complicated function of the parameters that enter the scale factor $a(t)$, or in mathematical terms, of the positions and residues of the remaining RSPs. If we denote the trace of ${\cal M}_{0\infty}$ by $T=-2 \cos 2\pi \sigma$, it is well known that $T^2$ determines the class of any $SL(2,\bC)$ matrix. Since in our case the trace is real, the possible classes are {\it elliptic} ($T^2 < 4$), parabolic ($T^2 =4$), and hyperbolic ($T^2> 4$). While the monodromy matrix  ${\cal M}_{0\infty}$ is basis-dependent, its trace is not. Specifically, the trace of ${\cal M}_{0\infty}$ in two different bases is
\be
\text{tr }
\begin{pmatrix}
\alpha & \beta \\
\beta^* & \alpha^*
\end{pmatrix}
\begin{pmatrix}
e^{-2\pi s\omega_f} & 0\\
0 & e^{2\pi s\omega_f} 
\end{pmatrix}
\begin{pmatrix}
\alpha^* & -\beta \\
-\beta^* & \alpha
\end{pmatrix}
\begin{pmatrix}
e^{2\pi s\omega_i} & 0 \\
0 & e^{-2\pi s\omega_i}
\end{pmatrix}= 
\text{tr } 
\begin{pmatrix}
-e^{2\pi i \sigma} & 0 \\
0 & -e^{-2\pi i \sigma}
\end{pmatrix}
\ee
which simplifies to
\be
 |\alpha_k|^2 \cosh 2\pi s(\omega_f-\omega_i)-|\beta_k|^2 \cosh 2\pi s(\omega_i+\omega_f)  =- \cos 2\pi \sigma 
\label{bogotrace}
\ee
Plugging (\ref{bogodet}) in (\ref{bogotrace}), after some elementary trigonometry we find
\be
|\beta_k|^2=\frac{\cos^2 \pi \sigma+\sinh^2 s\pi (\omega_f-\omega_i)}{\sinh 2\pi s\omega_i \, \sinh 2\pi s\omega_f}.
\label{bogogen}
\ee
Notice that in the most generic case $\sigma$ can be complex, but in our case where the ODE has real coefficients, and each basis is formed by a pair of complex conjugated solutions, $\sigma$ must be such that $|\beta_k|^2$ is real and positive. This gives three possibilities, that are closely related to the trace classification discussed above, but do not quite coincide. The reason is that $|\beta_k|^2$ depends on the trace, while the $SL(2,\bC)$ class depends on the square of the trace. If $-2< T<2$ we are in the elliptic case, or regime I, 
\be
|\beta_k|^2_I=\frac{\cos^2 \pi \sigma+\sinh^2 s\pi (\omega_f-\omega_i)}{\sinh 2\pi s\omega_i \, \sinh 2\pi s\omega_f}
\label{bogoI}
\ee
with $\sigma$ real. If $T<-2$, the conjugacy class is hyperbolic, $\sigma =i \Delta$ is purely imaginary and we call this regime II, 
\be 
|\beta_k|^2_{\text{II}}=\frac{\cosh^2 \pi \Delta+\sinh^2 \pi s(\omega_f-\omega_i)}{\sinh 2\pi s\omega_i \, \sinh 2\pi s\omega_f}.
\label{bogoII}
\ee
Finally, the trace can be $T>2$; we will not find this possibility realized in the examples that we study. If this case if also realized, we would call it regime III.

This is our first result. For any profile of the type (\ref{normalfuchsian}), the amount of particle production is of one of the previous forms, (\ref{bogoI}) or (\ref{bogoII}), where the dependence on the asymptotic values of the profile $\omega_i,\omega_f$ is cleanly separated from the intricacies of the profile, encoded in $\sigma$. Moreover, the two regimes have a different parametric dependence on $\theta$, oscillatory in the first case, and exponential in the second. This implies that the second regime can lead to enhanced particle production.

The specific formulas presented above are valid for the massless minimally coupled case. We can repeat the argument for the conformally coupled massive case, $m\neq 0$, $\xi=\frac{1}{6}$. The only difference comes from the different definition of the modes  - compare (\ref{confmode}) to (\ref{minmode}) - and it translates into a prefactor involving the scale factors. We arrive at
\be
|\beta_k|^2_I= \frac{a_f^2}{a_i^2} \frac{\cos^2 \pi \sigma+ \sinh ^2 \pi s (\omega_w-\omega_i)}{\sinh 2\pi s \omega_i\, \sinh 2\pi s \omega_f}
\ee
for the first regime. In the second regime, 
\be
|\beta_k|^2_{II}= \frac{a_f^2}{a_i^2} \frac{\cosh^2 \pi \Delta+ \sinh ^2 \pi s (\omega_f-\omega_i)}{\sinh 2\pi s \omega_i \, \sinh 2\pi s \omega_f}.
\ee
Let's discuss various features of this result. The parameter $s$ controls the width of the region where the profile changes. In the limit $s\rightarrow 0$, all these profiles collapse to a step function,
\be 
\omega(t)^2=\begin{cases}
\omega_f^2, & t>0 \\
\omega_i^2, & t<0 
\end{cases}
\ee
Since the problem is then very similar to the reflection of light between two media of different refractive index, we call this limit the Fresnel regime. In the Fresnel regime, the profile collapses to a step function erasing the details of the scale factor, and particle creation is given by
\be
|\beta_k|^2 =\left(\frac{a_f}{a_i}\right)^b \frac{(\omega_f-\omega_i)^2}{4 \omega_f \omega_i}
\label{fresnel}
\ee
where $b=0$ for the $m=\xi=0$ case, and $b=2$ for the massive conformally coupled scalar. While we don't have a proof, we think it is rather likely that for given $\omega_i, \omega_f$ this is the maximum amount of particle production.

For very large $|\vec k|$,  particle production is exponentially suppressed in the regime I,
\be  
|\beta_{\vec k}|^2_{\text{I}} \sim e^{-4\pi s\omega_i}.
\ee
This is precisely the result found in \cite{parker76} for a concrete example in the simplest case, a Fuchsian equation with three RSPs. On the other hand, in the regime II, if the condition $\Delta > s(\omega_f-\omega_i)$ is also satisfied, the similar formula is
\be 
|\beta_{\vec k}|^2_{\text{II}} \sim e^{2\pi (\Delta-s(\omega_f+\omega_i) )}.
\ee 
The discussion we have presented is valid for any Fuchsian ODE. On the other hand, it doesn't yield an explicit expression for $\sigma$ in terms of the details of the profile. We now turn to the simplest relevant examples of (\ref{normalfuchsian}), those coming from ODEs with 3 and 4 regular singular points.

\subsubsection{Hypergeometric profiles}
The first relevant example of the profiles (\ref{normalfuchsian}) appears when considering a mode function equation with precisely 3 regular singular points. As argued above, by a change of variables, any such  differential equation can be brought to the form (\ref{qform}), with 
\be
\omega^2(\eta)=\frac{\omega_f^2 e^{\frac{2 \eta }{s}}+\epsilon \, e^{\frac{\eta}{s}}+\omega_i^2}{(e^{\frac{\eta}{s}}+1)^2}.
\label{HGepstein}
\ee
The explicit example considered by Bernard and Duncan \cite{Bernard:1977pq} for a conformally coupled massive scalar field is the particular case $\epsilon = \omega_f^2+\omega_i^2$. In this case, the expansion profile simplifies to a hyperbolic tangent,
\be 
\omega^2(\eta)=\frac{\omega_f^2+\omega_i^2}{2}+\frac{\omega_f^2-\omega_i^2}{2} \tanh \frac{\eta}{2s}.
\label{bernard}
\ee
This is also the specific example considered by Parker \cite{parker76}, for a masless minimally coupled escalar. As already noted in \cite{parker76}, the generic hypergeometric profile in the form (\ref{HGepstein}) was discussed originally by Epstein \cite{epstein} in a different context. The new parameter $\epsilon$ introduces an asymmetry in the profile. As we vary $\epsilon$ keeping $\omega_{i,f}$ fixed, (\ref{HGepstein}) has a minimum for $\epsilon< 2\omega_i^2$, a maximum for $\epsilon>2 \omega_f^2$, and it is monotonous otherwise. 

\begin{figure}[h]
\centering
\includegraphics[scale=0.3]{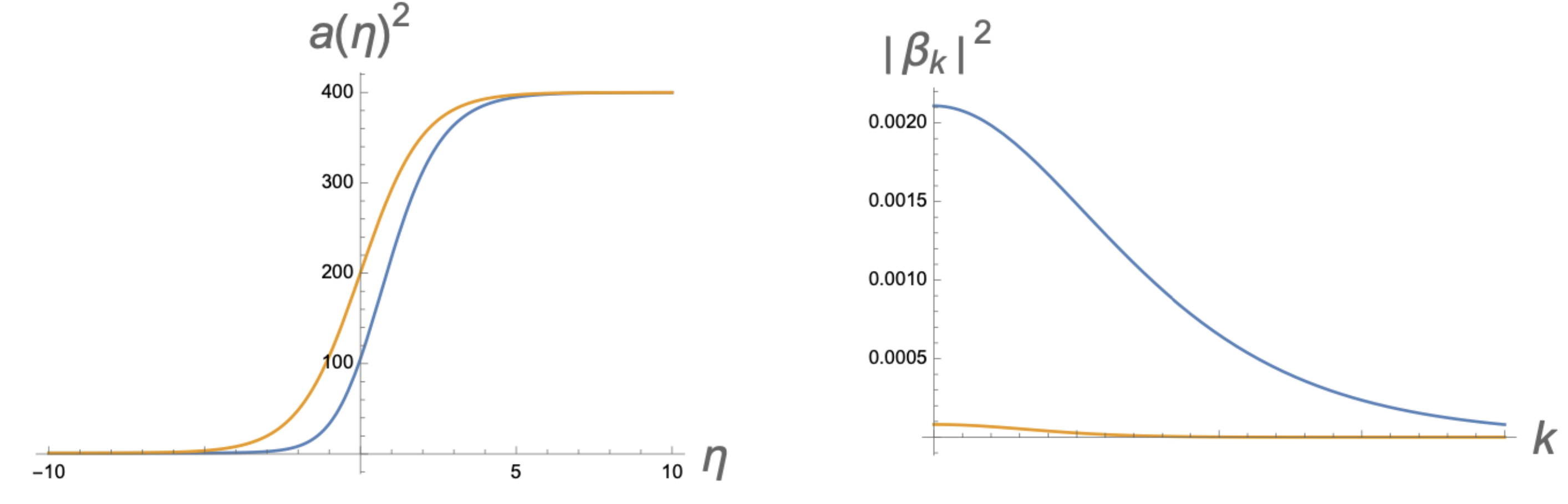}
\caption{Two examples of scale factors $a(\eta)$ in different regimes, and their respective production of massive conformally coupled scalar particles. The common values are $a_i=1$, $a_f=20$, $s=1$, $m=1$. The orange curves correspond to the symmetric profile (\ref{bernard}) with $\epsilon=a_i^2+a_f^2=401$ (regime I), while the blue curves correspond to $\epsilon=20$ (regime II). }
\label{scalesandbetas}
\end{figure}

Traditionally, the computation of particle production in this case is presented by first noticing that the relevant physical solutions involve hypergeometric functions, and then using the explicit connection formula for hypergeometric functions. In what follows, we present a shortcut, taking advantage of the previous discussion. In this simple case, the composite monodromy is just the monodromy around the third RSP. A short computation reveals that the indicial equation is
\be
r(r-1)+s^2 (\omega_f^2+\omega_i^2-\epsilon)=0
\ee
so in this case 
\be
\sigma =\sqrt{\frac{1}{4}+s^2 (\epsilon-\omega_i^2-\omega_f^2)}
\label{sigmahg}
\ee
The two regimes of production are distinguished by the sign of the quantity inside the square root of (\ref{sigmahg}): when it is positive, $\sigma$ is real, and we are in regime I of particle production, while when it is negative, $\sigma$ is purely imaginary, and we are in regime II. In figure \ref{scalesandbetas} we compare two scale factors belonging to different regimes, and their respective amounts of particle production.

\begin{figure}[h]
\centering
\includegraphics[scale=0.3]{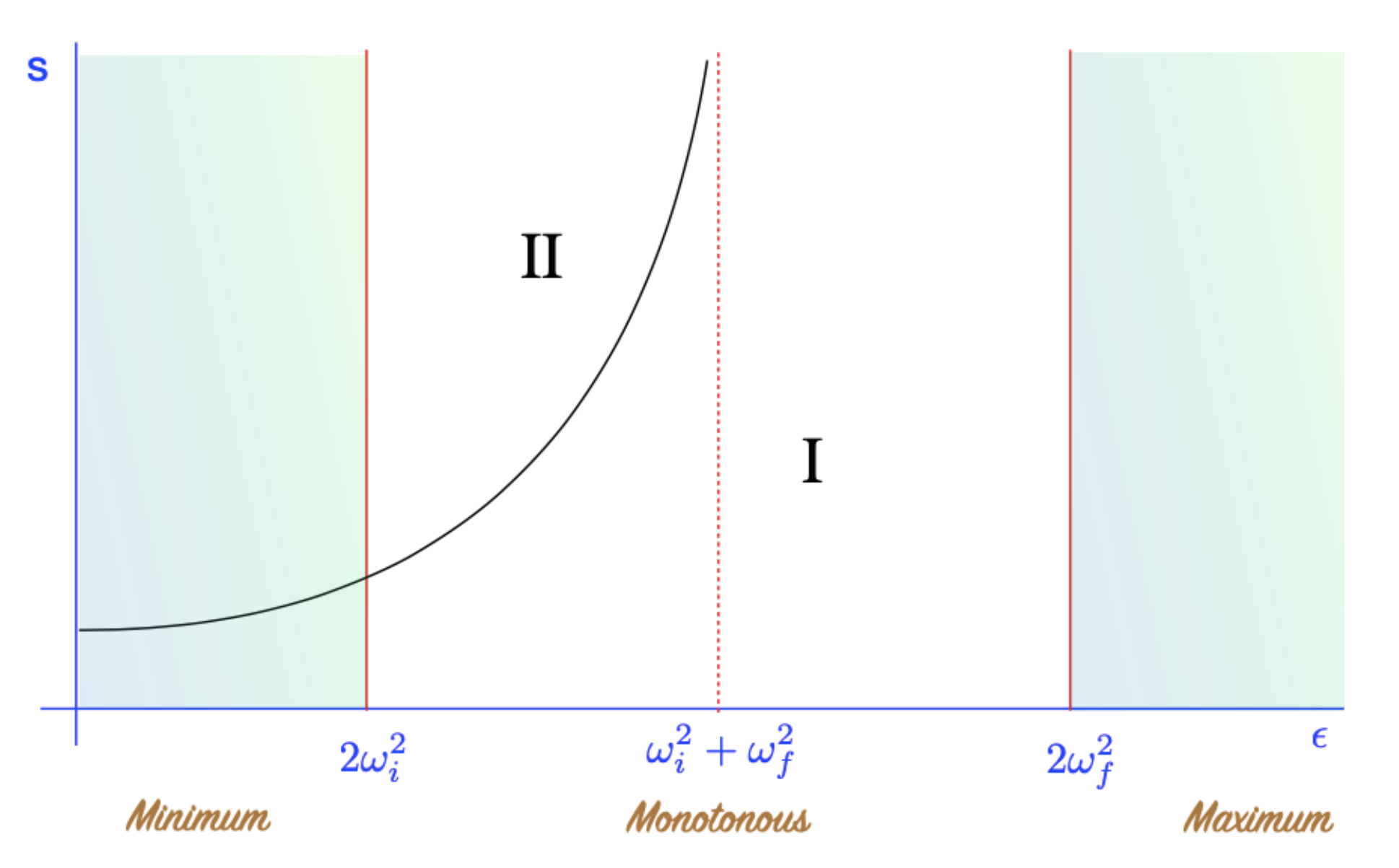}
\caption{The two regimes of particle production for hypergeometric profiles (\ref{HGepstein}). The examples discussed in \cite{Bernard:1977pq}  and in \cite{parker76} correspond to $\epsilon=\omega_i^2+\omega_f^2$ so they belong to regime I. }
\label{regimes}
\end{figure}

The boundary of both regimes thus takes place at $\epsilon_c=\omega_i^2+\omega_f^2-\frac{1}{4s^2}$. Notice that $\epsilon_c$ depends on $|\vec k|$ through $\omega_{i,f}$. The sign of $\epsilon_c$ is not constrained, so in principle the separation between these two regimes happens for arbitrarily small $s$. On the other hand, if we restrict to considering monotonous scale factors, the second regime only appears for $s\geq s_c$ with
\be
s_c=\frac{1}{2\sqrt{(\omega_f^2-\omega_i^2)}}
\ee
So for monotonous profiles, regime II only appears if the slope is not too steep. Moreover, it only appears when the asymmetry  $\epsilon-\omega_i^2-\omega_f^2$ in the scale factor is such that the expansion is delayed, not advanced. The various possibilities are summarized in figure \ref{regimes}.

Recall that the cases discussed in \cite{Bernard:1977pq}  and in \cite{parker76}  both correspond to $\epsilon=\omega_i^2+\omega_f^2$, so  according to (\ref{sigmahg}), $\sigma = \pm \frac{1}{2}$, and they belong to regime I. In the limit where the width of the expansion period $s$ gets smaller, all profiles tend to a step function. We now want to check that accordingly, the formula for particle production becomes universal. In (\ref{bogoI}), as $s$ tends to zero, $\cos^2 \pi \sigma$ is quartic in $s$, so it can be neglected compared to the $\sinh^2 \pi s (\omega_f-\omega_i)$ term. This implies that as $s$ tends to zero, $|\beta_{\vec k}|^2$ becomes independent of $\sigma$, approaching the Fresnel result, eq. (\ref{fresnel}). 

Finally, for this family of scale factors we can sharpen the general discussion about the $s\omega_{i,f} \gg 1$ regime, making it very precise. As explained above, in the regime I we have
\be  
|\beta_k|^2_{\text{I}} \sim e^{-4\pi s\omega_i}.
\ee
On the other hand, in the regime II, the condition $\Delta > s(\omega_f-\omega_i)$, boils down now to $\epsilon < 2 \omega_i \omega_f -\frac{1}{4s^2}$. When this condition is met, we have
\be 
|\beta_k|^2_{\text{II}} \sim e^{2\pi (\Delta- s(\omega_f+\omega_i))}.
\ee 
For monotonous profiles it is immediate to check that $\Delta- s(\omega_f+\omega_i)<0$ always holds, so there is still exponential suppression of particle creation. In summary, even inside regime II, the exponential suppression of large $s\omega$ particles presents two different parametric dependences as we vary the scale factor.

\subsubsection{Heun profiles}
\label{bosheun}
After having discussed mode function equations  with three regular singular points, we move to the next case: a mode function equation with four regular singular points, known as the Heun equation. This case presents qualitative new features that are common to all equations with four or more regular singular points. The corresponding scale factor is
\be
\omega(\eta)^2 =\frac{\omega_f^2 e^{4\eta/s}+ \epsilon_3 \, e^{3\eta/s}+ \epsilon_2 \, e^{2\eta/s} +\epsilon_1 \, e^{\eta/s}+ t^2 \omega_i^2}{(e^{\eta/s}+1)^2 (e^{\eta/s}+t)^2}.
\label{epsheun}
\ee
Following our general discussion, particle production in FLRW spacetimes with this scale factor is given either by (\ref{bogoI}) or (\ref{bogoII}). Our goal is again to find the composite monodromy exponent $\sigma$ as a function of the parameters of the scale factor (\ref{epsheun}), and delimitate the boundary between regimes I and II of particle production in this case.

There has been recent progress in deriving the connection formula for the Heun equation \cite{Bonelli:2022ten, Lisovyy:2022flm}, but since we are not interested in the full connection matrix, we will only rely on a subset of these results. In those works it is customary to write the Heun equation as \cite{Lisovyy:2022flm}
\be
\begin{split}
\frac{d^2\Psi}{dz^2}+ 
\left[\frac{\frac{1}{4}-\theta_0^2}{z^2}+
\frac{\frac{1}{4}-\theta_1^2}{(z-1)^2}+
\frac{\frac{1}{4}-\theta_t^2}{(z-t)^2}+
\frac{\theta_0^2+\theta_1^2+\theta_t^2-\theta_\infty^2-\frac{1}{2}}{z(z-1)} \right. \\
\left. +  \frac{(t-1)(-\nu^2+\theta_t^2+\theta_1^2-\frac{1}{4})}{z(z-1)(z-t)}\right] \Psi=0
\end{split}
\label{heunlisovyy}
\ee
M\"obius transformations can fix the locations of up to three RSPs. In this case, we place three RSPs at $0,1,\infty$ and the fourth is at $t$. Written in this way, the four RSPs are at $z=0,1,t,\infty$ and the respective local exponents are $\frac{1}{2}\pm \theta_0$, $\frac{1}{2}\pm \theta_1$, $\frac{1}{2}\pm \theta_t$ and $-\frac{1}{2}\pm \theta_\infty$. 

Notice that there is a parameter $\nu$ in the differential equation (\ref{heunlisovyy}) that does not appear in any of the local exponents. This implies that knowledge of the local monodromies around individual RSPs does not determine the equation.
$\nu$ is called an accessory parameter, and determining the dependence on $\nu$ of the composite monodromy is rather non trivial.

Our first task is to relate the parameters in (\ref{epsheun}) and (\ref{heunlisovyy}). A long computation yields
\be 
\begin{split}
& \theta_0^2 = -s^2 \omega_i^2, \\
& \theta_\infty^2=-s^2 \omega_f^2, \\
& \theta_1^2 =\frac{1}{4}-s^2 \frac{\omega_f^2+t^2 \omega_i^2 -\epsilon_3+\epsilon_2-\epsilon_1}{(t-1)^2}, \\
& \theta_t^2 = \frac{1}{4}-s^2 \frac{t^3 \omega_f^2+t \omega_i^2-t^2 \epsilon_3+t\epsilon_2-\epsilon_1}{(t-1)^2}, \\
& \nu^2= \frac{1}{4}+s^2 \omega_i^2 -s^2 \omega_f^2 +s^2 \frac{2t\omega_f^2+(4t^2-2t^3)\omega_i^2-(t+1)\epsilon_3+2\epsilon_2+(t-3)\epsilon_1}{(t-1)^3}
\end{split}
\label{bosoheun}
\ee
It is worth mentioning that in much of the literature devoted to the connection problem \cite{Lisovyy:2022flm}, $\nu^2$ is a completely free parameter, and therefore manifestly independent on the coordinate $t$ of the fourth RSP. On the other hand, in this work, the scale factor $a(\eta)$ determines completely the mode function equation, and since (\ref{epsheun}) depends explicitly on $t$, $\nu^2$ will in general depend on $t$ too.

The relations (\ref{bosoheun}) can be intimidating, so it is reassuring to perform a check on them. When
\be
\begin{split}
&\epsilon_3 = 2t \omega_f^2 +\epsilon, \\
&\epsilon_2 =t^2 \omega_f^2+\omega_i^2+2t\epsilon, \\
&\epsilon_1 = 2t \omega_i^2+t^2 \epsilon
\end{split}
\ee
the profiles (\ref{epsheun}) reduce to the hypergeometric one, (\ref{HGepstein}). For these particular values, (\ref{bosoheun}) simplifies to $ \theta_0^2 = -s^2 \omega_i^2,  \theta_\infty^2=-s^2 \omega_f^2$ and
\be
\theta_t^2=\frac{1}{4}, \hspace{1cm} \nu^2=\theta_1^2=\frac{1}{4}+s^2 (\epsilon-\omega_i^2-\omega_f^2).
\ee
These are indeed the expected values in the hypergeometric limit: in this limit the RSP at $z=t$ in (\ref{heunlisovyy}) must disappear, and $\theta_t^2=\frac{1}{4}$ guarantees that;  on the other hand, the composite monodromy exponent $\sigma$ is given by the monodromy around the third RSP, $\sigma=\theta_1$, and indeed this matches the hypergeometric result (\ref{sigmahg}).

The next task is to write the exponent of the composite monodromy, $\sigma$, in terms of the parameters of the Heun equation. We will explain how this can be accomplished when the Heun equation is written in the form (\ref{heunlisovyy}). As it turns out, this is a problem that has appeared a number of times in various physical setups, so it can be solved using different approaches, although we are not aware of a closed form expression. Here we will exploit the relation of the Heun equation and the classical limit of the Virasoro conformal blocks of 2D conformal field theory, following \cite{Litvinov:2013sxa, Lisovyy:2022flm}.



In two-dimensional conformal field theory, Virasoro conformal blocks are the building blocks of n-point functions. In particular, the conformal block associated to the sphere 4-point function depends on 4 scaling dimensions $\Delta_i$, the four positions of the punctures $z_i$ and the momentum $P$ of the intermediate state. By a M\"obius transformation, we can bring the position of three of the punctures to $0,1,\infty$, and we denote the resulting conformal block by $F(0,1,t,\infty)$. In the limit of large central charge $c\rightarrow \infty$, the Virasoro conformal block is conjectured \cite{zamo} to exponentiate
\be 
F(\Delta_i) \sim e^{\frac{c}{6} W(\delta_i)}
\ee
where W is called the classical conformal block, that depends on the classical dimensions $\delta_a=\frac{1}{4}-\theta_a^2$. The classical conformal block admits a perturbative expansion in $1/t$ of the form
\be 
W(t)=(-\delta_1+\delta_\sigma-\delta_t) \ln t +\sum_{k=1}^\infty W_k t^{-k} 
\ee
Then, if we write a perturbative expansion for
\be
\sigma (t)=\sum_{k=0}^\infty \sigma_k t^{-k}
\ee
one has \cite{Lisovyy:2022flm} 
\be
\sigma(t)^2 =\nu^2-\sum_{k=1}^\infty k W_k t^{-k}
\label{sigmafromw}
\ee
so we can find the coefficients $\sigma_k$ recursively, using the known perturbative expansion of the classical conformal block W \cite{zamo}. Again, in our case, $\nu^2$ will depend on $t$, so it has to be expanded in $1/t$ in (\ref{sigmafromw}). At large $t$, we find
\be
\sigma^2 = \lim_{t\to \infty} \nu^2 = \frac{1}{4}-s^2 (\omega_i^2+\omega_f^2) 
\ee
We thus find that at large $t$, the boundary between regime I and regime II is at $\frac{1}{4}=s^2 (\omega_i^2+\omega_f^2)$. When $\sigma^2$ is positive, we are in regime I, and when $\sigma^2$ is negative, we are in regime II. Using the known perturbative expansion of $W(t)$, one could compute $1/t$ corrections to this boundary.


\section{Spin 1/2 particle production}
The study of cosmological production of Dirac fermions is conceptually similar to the scalar case, but technically more complicated, as it requires dealing with the generalization of Dirac's equation to curved spacetimes \cite{Parker:1971pt}. In the case of flat FLRW metrics, the mode equation turns out to be \cite{Duncan:1977fc}
\be 
\frac{d^2 \phi_k}{d \eta^2} +\left (|\vec k|^2+m^2 a^2(\eta) + m \gamma^0 \dot a(\eta) \right) \phi_k=0
\ee
It is convenient to split it into two equations, corresponding to the eigenspaces of $\gamma^0$ \cite{Duncan:1977fc},
\be
\frac{d^2 \phi_k}{d \eta^2} +\left (|\vec k|^2+m^2 a^2(\eta) \pm i m \dot a(\eta) \right) \phi_k=0
\label{fermmod}
\ee
These are two Schr\"odinger equations, but with a complex potential. Each of this two equations has a two-dimensional vector space of solutions. However, contrary to what happened in the scalar case, since now the potential is manifestly complex, the complex conjugate of a solution is not a solution of the same equation; rather it is a solution of the other equation. 

As for the massive scalar case, let's denote the asymptotic values of the frequencies by
\be
\omega_i(\vec k)=\sqrt{|\vec k|^2 + m^2 a^2_i}, \hspace{1cm} \omega_f(\vec k)=\sqrt{|\vec k|^2 + m^2 a^2_f}
\ee
Then, in each $\gamma^0$-eigenspace there will be a solution that tends to $e^{-\omega_i(\vec k)t}$ as $t\to -\infty$. Let's denote these two solutions by $\phi_i^\pm$. Similarly, $\phi_f^\pm$ denote the two solutions that asymptote to $e^{-\omega_f(\vec k)t}$, one in each eigenspace of $\gamma^0$. Now, for each of these four solutions, their complex conjugates are solutions of the equation in the other eigenspace. The upshot of this discussion is that the two bases for the + eigenspace are $\phi_i^+, \phi_i^{-*}$ and $\phi_f^+,\phi_f^{-*}$. They are related by a Bogoliubov transformation,
\be
\begin{pmatrix}
\phi_i^+ \\
 \phi_i^{-*}
\end{pmatrix}
= 
\begin{pmatrix}
\alpha^+ & \beta^+ \\
\beta^{-*} & \alpha^{-*} 
\end{pmatrix}
\begin{pmatrix}
\phi_f^+ \\
\phi_f^{-*}
\end{pmatrix}
\label{fermbogo}
\ee
Similarly, the two bases for the second eigenspace are $\phi_i^-, \phi_i^{+*}$ and $\phi_f^-,\phi_f^{+*}$, which are related by the complex conjugate of the Bogoliubov transformation above. These Bogoliubov coefficients are not independent, they satisfy \cite{Duncan:1977fc},
\be 
\begin{split}
& \frac{\alpha^+}{\alpha^-}  = \frac{\omega_i-m a_i}{\omega_f-m a_f}=\frac{\omega_f+m a_f}{\omega_i+m a_i}, \\
& \frac{\beta^+}{\beta^-}  = -\frac{\omega_i-m a_i}{\omega_f+m a_f}=-\frac{\omega_f-m a_f}{\omega_i+m a_i}, \\
& \alpha^-\alpha^{+*}-\beta^- \beta^{+*}  = \frac{\omega_i}{\omega_f}
\end{split}
\label{duncanrel}
\ee
Note that in the fermionic case we are not using the plane wave normalization, to ease the comparison of our formulas with \cite{Duncan:1977fc}. As in the scalar case, we are going to restrict to mode function equations that are Fuchsian, {\it i.e.} only have regular singular points. Before we discuss particle production, let's pause to elucidate what scale factors $a(\eta)$ satisfy this requirement. Now, according to (\ref{fermmod}), we must demand that
\be 
m^2 a(\eta)^2 \pm i \, m \, \dot a(\eta) =\frac{\epsilon_{2n} e^{2n \eta/s} +\epsilon_{2n-1} e^{(2n-1)\eta/s}+\dots +\epsilon_0}{(e^{\eta/s}+c_1)^2 \dots (e^{\eta/s}+c_n)^2}
\ee
If we impose that $a(\eta)$ tends to constant values as $\eta \to \pm \infty$, then $\dot a \to 0$ in those limits. This fixes $\epsilon_{2n}$ and $\epsilon_0$ in the equation above,
\be 
m^2 a(\eta)^2 \pm i \, m \, \dot a(\eta) =\frac{m^2 a_f^2 e^{2n \eta/s} +\epsilon_{2n-1} e^{(2n-1)\eta/s}+\dots +(c_1\dots c_n)^2 m^2 a_i^2}{(e^{\eta/s}+c_1)^2 \dots (e^{\eta/s}+c_n)^2}.
\label{epsfermion}
\ee
The remaining $\epsilon$ coefficients in (\ref{epsfermion}) must now be complex, while they were real in the bosonic case. Taking the imaginary part of the eq. (\ref{epsfermion}), recalling that $a(\eta)$ is real and granting that the $c_i$ are real, we have an equation for $\dot a(\eta)$. The solution that is consistent with (\ref{epsfermion}) is
\be
a(\eta)=a_f-\frac{A_1}{e^{\eta/s}+c_1}-\dots-\frac{A_n}{e^{\eta/s}+c_n},
\label{fermscale}
\ee
with the constraint that $m^2 a(\eta)^2\pm i m \dot a \to m^2 a_i^2$ as $\eta \to -\infty$. Moreover, the constants appearing in (\ref{fermscale}) will have to satisfy additional constraints if we want to restrict to monotonous scale factors. The scale factors (\ref{fermscale}) are a subset of the scale factors allowed in the scalar case. As discussed below, the explicit example in \cite{Duncan:1977fc} is a particular case of this family of scale factors.

Let's now give a general formula for Dirac fermion production in those spacetimes. We can repeat the argument we presented in the scalar case, and consider for each of the two equations in (\ref{fermmod}) a composite monodromy encircling the two regular singular points in the asymptotic past and the asymptotic future. The reasoning of the scalar case doesn't immediately apply now, since the components of the Bogoliubov transformation (\ref{fermbogo}) are not complex conjugate of each other as before. Luckily, the relations (\ref{duncanrel}) between the Bogoliubov coefficients in the two spaces still allow to derive expressions for $|\beta^{\pm}_{\vec k}|^2$. Contrary to the scalar case, now formulas are neater if we take the convention that the eigenvalues of the composite monodromy matrix are $\{ e^{2\pi i \sigma}, e^{-2\pi i \sigma}\}$. We find 
\be
\begin{split}
|\beta^+_k|^2 & =\frac{\omega_i}{\omega_f}\frac{\omega_f-m a_f}{\omega_i-m a_i}
\frac{\cos 2\pi \sigma - \cosh 2 \pi (\omega_f-\omega_i)}{\cosh 2\pi (\omega_f+\omega_i)-\cosh 2\pi (\omega_f-\omega_i)}, \\
|\beta^-_k|^2 & =\left( \frac{\omega_f+ma_f}{\omega_i-ma_i}\right)^2 |\beta_k^+|^2. 
\end{split}
\ee
Since $|\beta^+_k|^2$ and $|\beta_k^-|^2$ have to be real and positive, now $\sigma$ can't be real. Instead $\sigma$ has to be purely imaginary $\sigma = i \Delta$. So we learn that contrary to what happens in the scalar case, there is only one regime of particle production,
\be
|\beta^+_k|^2 =\frac{\omega_i}{\omega_f}\frac{\omega_f-m a_f}{\omega_i-m a_i}
\frac{\cosh 2\pi \Delta - \cosh 2 \pi (\omega_f-\omega_i)}{\cosh 2\pi (\omega_f+\omega_i)-\cosh 2\pi (\omega_f-\omega_i)}
\label{fermbeta}
\ee
To conclude, let's illustrate these formulas for fermion production with concrete examples. The simplest scale factor to consider in the family (\ref{fermscale}) is 
\be
a(\eta)=a_f-\frac{a_f-a_i}{e^{\eta/s}+1}=\frac{a_i+a_f}{2}+\frac{a_f-a_i}{2} \tanh \frac{\eta}{2s}
\label{duncan}
\ee
It corresponds to a mode function equation with three regular singular points. This is precisely the scale factor discussed in \cite{Duncan:1977fc} as an example of cosmological fermion production. Note that this scale factor is {\it not} the same as (\ref{bernard}), which corresponds to the explicit example in \cite{Bernard:1977pq} for a massive scalar field. A short computation reveals that in this case $|\beta_{\vec k}^+|^2$ is of the form (\ref{fermbeta}) with $\Delta= s m (a_f-a_i)$. Of course this value agrees with the one obtained in \cite{Duncan:1977fc} from the linear relations among the explicit solutions. Again, our computation is much shorter and it doesn't rely on the solutions to the mode function equations. Note that for (\ref{fermbeta}) to make sense, it must be the case that $m(a_f-a_i)> (\omega_f-\omega_i)$; it is immediate to check that this is the case.

The next example of scale factors in (\ref{fermscale}) is
\be
a(\eta)=a_f-\frac{A_1}{e^{\eta/s}+1}-\frac{A_2}{e^{\eta/s}+t}
\label{heunferm}
\ee
with the constraint $a_f-A_1-\frac{A_2}{t}=a_i$. It corresponds to a mode function equation with four regular singular points. A tedious computation shows that the mode function equations can be brought to the form (\ref{heunlisovyy}), with the following identifications
\be 
\begin{split}
& \theta_0^2  =-s^2k^2-s^2m^2 a_i^2, \\
& \theta_\infty^2  =-s^2 k^2-s^2 m^2 a_f^2, \\
& \theta_1^2  = \left(i ms A_1\pm \frac{1}{2}\right)^2, \\
& \theta_t^2  = \left(i ms\frac{A_2}{t} \pm \frac{1}{2} \right)^2, \\
& \nu^2 =(imsA_1\pm\frac{1}{2})^2-m^2s^2 \left(\frac{A^2_2}{t^2}+2a_i {A_2}{t}+2 \frac{A_1 A_2}{t-1}\right).
\end{split}
\ee
As a check of this result, note that in the $A_2\to 0$ limit, we recover the hypergeometric result. This is indeed the case, since then $A_1=a_f-a_i$, $\delta_t=0$ and the composite monodromy parameter is $\sigma=\nu-1/2$, so $\cos 2\pi  \sigma=\cosh ms (a_f-a_i)$.
\acknowledgments
We would like to thank Jaume Garriga for conversations on cosmological particle production. Research supported by  the State Agency for Research of the Spanish Ministry of Science and Innovation through the ``Unit of Excellence Mar\'ia de Maeztu 2020-2023'' award to the Institute of Cosmos Sciences (CEX2019-000918-M), by grants PID2019-105614GB-C22 and  PID2022-136224NB-C22, funded by MCIN/AEI/ 10.13039/501100011033/FEDER, UE and PID2019-105614GB-C22, and by AGAUR, grant 2021 SGR 00872. The research of J. R.-P. is supported by a grant from the Spanish Ministry of Science and Innovation for collaboration with University departments, 2023 COLAB 00363.

\appendix

\section{Fuchsian differential equations}
In this appendix we collect some very basic mathematical facts needed in the main body of the text. A traditional source on ordinary differential equations is \cite{ince}. A more recent book is \cite{haraoka}. Consider a second order homogeneous linear ordinary differential equation
\be
y''(z)+p_1(z) y'(z)+p_2(z) y(z)=0,
\ee
with $z\in \bC \bP^1$ (the Riemann sphere). We classify points $z$ according the behaviour of $p_i(z)$ as follows. $z_0$ is an {\it ordinary point} if $p_i(z)$ are well defined at $z_0$. We say that $z_0$ is a {\it regular singular point} (RSP) if $p_1(z)$ or $p_2(z)$ is not defined at $z_0$, but the limits of
\be
(z-z_0) p_1(z), \hspace{1cm} (z-z_0)^2 p_2(z)
\ee
as $z\rightarrow z_0$ exist and are finite. If $z_0$ is neither ordinary nor a RSP, we say that $z_0$ is an {\it irregular singular point}. When counting the number of singular points of an ODE, it is mandatory to check whether $z=\infty$ is a singular point or not. For this purpose, the standard procedure is to perform the change of variable $t=1/z$, and then apply the criterion above to $t=0$.

An ordinary differential equation such that all its singular points are regular is called a Fuchsian differential equation. The most general second order Fuchsian differential equation is
\be 
\frac{d^2 \Psi(z)}{dz^2}+\left(\sum_{k=1}^n \frac{A_k}{z-z_k}\right) \frac{d\Psi(z)}{dz}+
\sum_{k=1}^n \left( \frac{B_k}{(z-z_k)^2}+\frac{C_k}{z-z_k} \right) \Psi(z)=0
\label{appfuchs}
\ee
This equation has RSPs at $z_1,\dots,z_n$. Moreover $z=\infty$ is also a RSP unless $\sum_k A_k=2$ and $ \sum_k (B_k+z_k C_k)=0$.

\subsection{Solutions at regular singular points}
The space of solutions of (\ref{appfuchs}) is two-dimensional. The Frobenius method provides a basis of solutions for each RSP.
If $z_0$ is a RSP, define 
\be
a_i =\text{lim}_{z\rightarrow z_0} (z-z_0)^i \, p_i (z)\hspace{1cm} i=1,2
\ee
and the {\it indicial equation}
\be
r(r-1)+a_1r+a_2=0
\ee
The roots $r_i$ of the indicial equation for $z_0$ are the {\it local exponents} of the RSP. For RSPs, we are guaranteed at least the existence of a solution of the form
\be
y(z)=\sum_{k=0}^\infty a_k (z-z_0)^{k+r}
\label{appfrobe}
\ee
for some root $r$ of the indicial equation (Frobenius-type solution). The generic situation is when $r_1,r_2$ do not differ by an integer. In this case, a second solution exists with the same form, for the other root. If the roots of the indicial equation $r_1,r_2$ differ by an integer, the second solution is more complicated, but we will not discuss it here. The radius of convergence of the power series above is at least until the nearest RSP (Fuchs theorem).

The solutions (\ref{appfrobe}) are multivalued for non-integer $r$. If we analytically continue them along a small loop around the regular singular point, the original basis changes to a new basis. The matrix relating these two bases is the {\it local monodromy matrix}. Its eigenvalues are $e^{2\pi i r_i}$, where $r_i$ are the roots of the corresponding indicial equation.

\subsection{Operations on Fuchsian equations}
There are two interesting operations that transform a Fuchsian equation into another Fuchsian equation: 

1) M\"obius transformations of $z$
\be 
z\rightarrow \frac{az+b}{cz+d}
\ee
change the positions of up to three RSPs, but keep the roots of the indicial equations. 

2) An index transformation
\be 
\Psi(z)=(z-z_1)^{a_1}\dots (z-z_n)^{a_n} \Phi(z)
\ee 
keeps the position of the RSPs but changes the local exponents as follows:
\be 
r_{k,i} \rightarrow r_{k,i}-a_k, \hspace{1cm} r_{\infty,i}\rightarrow r_{\infty,i}+\sum_k a_k \, .
\ee

\subsection{Global features}
So far, we have discussed local properties of the solutions, near a regular singular point. In the study of cosmological particle production, it is crucial to go beyond local properties, and study global properties of the space of solutions.

A first question is the {\it connection problem} \cite{schafke}: given two bases at different RSPs, what is the change of basis relating them? A second type of global questions are {\it composite monodromies}: what is the monodromy matrix associated to a path encircling more than one regular singular point? When the differential equation has accessory parameters, {\it i.e.}, parameters that don't appear in elementary monodromies, this becomes a rather non-trivial question.

\subsection{Examples}
To conclude this brief overview, let's list the simplest second order Fuchsian ODEs, according to the number of regular singular points. 

\underline{1 RSP}. The ODE
\be
y''=0
\ee
has a RSP at $z=\infty$. Any other second order Fuchsian  ODE with just one RSP can be transformed into this one by a change of variables. We change variables $t=1/z$, and now the RSP is at $t=0$. The indicial equation for $t=0$ is $r^2+r=0$. The roots are $r=0,-1$. Thus the eigenvalues of the monodromy matrix are $1,1$. The monodromy matrix is the identity, ${\cal M}_\infty=1$. The general solution is $y(z)=c_1 z+c_2$.

\underline{2 RSP}. The Euler differential equation
\be
z^2 y''+a z y'+by=0
\ee
with $a,b$ constants has RSPs at $z=0,\infty$. Any second order Fuchsian ODE with just two RSPs can be transformed into this one by a change of variables. The indicial equation is $r^2+(a-1)r+b=0$. The local exponents are the two roots $r_{1,2}$ of this equation. The second RSP is at $z=\infty$. After changing variables $t=1/z$, the corresponding indicial equation is $r^2+(1-a)r+b=0$. The roots of this equation are minus the roots of the previous one. Thus, the eigenvalues of ${\cal M}_\infty$ are inverse of those of ${\cal M}_0$. If they are diagonal in the same basis, this proves ${\cal M}_0 {\cal M}_\infty =1$. The solutions around $z=0$ are
\be
y(z)=c_1 z^{r_1}+c_2 z^{r_2}
\ee

\underline{3 RSP}. The most general second order Fuchsian ODE with three RSPs is called Riemann's differential equation. By a change of variables, any Riemann differential equation can be brought to a form where the singularities are at $z=0,1,\infty$. In this case, the equation is called the hypergeometric differential equation
\be
z(1-z) \, y''(z) +[c-(a+b+1)z]\, y'(z)- a b \,y(z)=0
\label{hyperapp}
\ee
where $a,b,c$ are constants.  One can check that for the hypergeometric differential equation (\ref{hyperapp}), the local exponents at $0,1,\infty$ are $(0,1-c)$, $(0,c-a-b)$,$(a,b)$ respectively. For generic values of $a,b,c$ we can write two solutions of the hypergeometric equation (\ref{hyperapp}) in terms of the hypergeometric function $_2 F_1(a,b,c;z)$. Near $z=0$ two solutions are 
\be
F(a,b,c,z), \hspace{1cm} z^{1-c} F(1+a-c,1+b-c,2-c,z)
\ee
Near $z=1$,
\be
F(a,b,1+a+b-c,1-z),\hspace{1cm} (1-z)^{c-a-b} F(c-a,c-b,1+c-a-b,1-z)
\ee
Near $z=\infty$,
\be
z^{-a} F(a,1+a-c,1+a-b,z^{-1}), \hspace{1cm} z^{-b} F(b,1+b-c,1+b-a,z^{-1}) .
\ee

\end{document}